\documentclass{PoS}
\usepackage{epsf}

\newcommand{\third}{\mbox{\small $\frac{1}{3}$}}         
\newcommand{\twothird}{\mbox{\small $\frac{2}{3}$}}      

\def\lsim{\mathrel{\rlap{\lower4pt\hbox{\hskip1pt$\sim$}}
    \raise1pt\hbox{$<$}}}                
\def\gsim{\mathrel{\rlap{\lower4pt\hbox{\hskip1pt$\sim$}}
    \raise1pt\hbox{$>$}}}                
\title{
\vspace*{-2cm}
\begin{minipage}{\textwidth}
\begin{flushright}
\texttt{\footnotesize
PoS(LAT2009)102    \\%
DESY 09-172        \\%
Edinburgh 2009/15  \\%
Liverpool LTH 848  \\%
}
\end{flushright}
\end{minipage}\\[15pt]
\vspace*{+2cm}
       Results from 2+1 flavours of SLiNC fermions}

\ShortTitle{Results from 2+1 flavours of SLiNC fermions}

\author{W.~Bietenholz$^a$,
        V.~Bornyakov$^b$,
        N.~Cundy$^c$,
        M.~G\"ockeler$^c$,
        \speaker{R.~Horsley}$^{\,d}$,
        A.~D. Kennedy$^d$,
        Y.~Nakamura$^c$,
        H.~Perlt$^e$,
        D.~Pleiter$^f$,
        P.~E.~L.~Rakow$^g$,
        A.~Sch\"afer$^c$,
        G.~Schierholz$^{ch}$,
        A.~Schiller$^e$,
        H.~St\"uben$^i$
        and J.~M.~Zanotti$^d$ \\
        \llap{$^a$} Instituto de Ciencias Nucleares,
                    Universidad Aut\'{o}noma de M\'{e}xico,
                    A.P. 70-543, C.P. 04510 Distrito Federal, Mexico \\
        \llap{$^b$} Institute for High Energy Physics,
                    142281 Protovino, Russia and \\
                    Institute of Theoretical and Experimental Physics,
                    117259 Moscow, Russia \\
        \llap{$^c$} Institut f\"ur Theoretische Physik,
                    Universit\"at Regensburg,
                    93040 Regensburg, Germany \\
        \llap{$^d$} School of Physics and Astronomy,
                    University of Edinburgh,
                    Edinburgh EH9 3JZ, UK \\
        \llap{$^e$} Institut f\"ur Theoretische Physik,
                    Universit\"at Leipzig,
                    04109 Leipzig, Germany \\
        \llap{$^f$} John von Neumann Institute NIC / DESY Zeuthen,
                    15738 Zeuthen, Germany \\
        \llap{$^g$} Theoretical Physics Division,
                    Department of Mathematical Sciences,
                    University of Liverpool,
                    Liverpool L69 3BX, UK \\
        \llap{$^h$} Deutsches Elektronen-Synchrotron DESY,
                    22603 Hamburg, Germany \\
        \llap{$^i$} Konrad-Zuse-Zentrum f\"ur Informationstechnik Berlin,
                    14195 Berlin, Germany \\
        E-mail: \email{rhorsley@ph.ed.ac.uk} }

\author{QCDSF--UKQCD Collaborations}

\abstract{
QCD results are presented for a 2+1 flavour fermion clover action
(which we call the SLiNC action). A method of tuning the quark masses
to their physical values is discussed. In this method the singlet quark
mass is kept fixed, which solves the problem of different
renormalisations (for singlet and non-singlet quark masses)
occuring for non-chirally invariant lattice fermions.
This procedure enables a wide range of quark masses to be probed,
including the case with a heavy up-down quark mass and light
strange quark mass. Preliminary results show the correct splittings
for the baryon (octet and) decuplet spectrum.}

\FullConference{The XXVII International Symposium on Lattice Field Theory\\
		 July 26-31, 2009\\
		 Peking University, Beijing, China}

\begin{document}


\section{Introduction}


There has been a steady progression of lattice results from a quenched sea
to two-flavour and more recently $2+1$ flavour sea in an attempt
to provide a more complete and quantitative description of hadronic phenomena.
(By $2+1$ flavours we mean here $2$ mass degenerate up-down,
$m_l$, quarks and one strange, $m_s$, quark.) In this talk we shall
consider a $2+1$ flavour clover action and discuss some ways of
approaching in the $m_l$--$m_s$ plane the physical point $(m_l^*, m_s^*)$,
where the natural starting point for these paths is an $SU_F(3)$
flavour symmetric point $m_l = m_s = m_{sym}^{(0)}$. For clover
(i.e.\ non-chiral) fermions a problem arises because the measured
pseudoscalar masses are proportional to the {\it renormalised} quark masses.
The singlet, $S$, and nonsinglet, $NS$ quark masses renormalise differently
which means that the relation to the bare quark masses and hence
$\kappa$, which is the adjustable simulation parameter,
is more complicated \cite{gockeler04a}. Choosing the path
such that the singlet quark mass is kept fixed provides an elegant
solution to this problem. This procedure also has the
advantage that it enables a wide range of quark masses to be probed
(including the mass of the strange quark) and is thus particularly useful
for strange quark physics and $SU_F(3)$--chiral perturbation theory,
as the kaon mass is never larger than its physical value.
(Note that this path choice although favourable for clover--type
fermions is not restricted to them.) Indeed this includes the
case with a heavy light quark mass and light strange quark mass.
As a `proof of concept' preliminary results given here show the
correct splittings for the baryon (octet and) decuplet spectrum.

\begin{figure}[htb]
   \hspace{1.25in}
   \epsfxsize=7.00cm
      \epsfbox{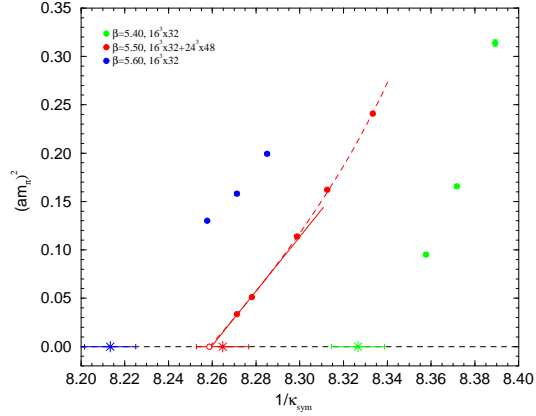}
   \caption{Results for the pseudoscalar mass against $1/\kappa_{sym}$
            for $\beta = 5.40$, $5.50$, $5.60$. For $\beta = 5.50$ a
            linear fit (line) using the lightest three masses is
            made. For comparison a quadratic fit (dashed line) is also
            made to all the masses. The stars represent the SF
            determination of $\kappa_{sym;c}$; the open circle
            is the result from the linear fit.}
\label{b5p40+b5p50+b5p60_amps2_ookl_lat09}
\end{figure}

The particular clover action used here has a single iterated
mild stout smearing for the hopping terms together
with thin links for the clover term (this is in an attempt
to ensure that the fermion matrix does not become too extended).
Together with the (tree level) Symanzik improved gluon action this
constitutes the Stout Link Non-perturbative Clover or SLiNC action
for which the clover coefficient, $c_{sw}$, has recently been
non-perturbatively (NP) determined, \cite{cundy09a}, using the
Schr\"odinger Functional, or SF, formalism. Further details
about the action may be found in this reference. Simulations
have been performed using HMC with mass preconditioning
for $2$ mass-degenerate flavours and the rational approximation for
the $1$-flavour. Two programmes were used, a Fortran programme,
\cite{gockeler07a}, and also the Chroma programme, \cite{edwards04a}.
Quark mass degenerate runs (denoted by the subscript `$sym$')
on $16^3\times 32$ lattices followed by $24^3\times 48$ lattices
have located a suitable $\beta$-range, see
Fig.~\ref{b5p40+b5p50+b5p60_amps2_ookl_lat09}.
We see that there is good agreement between SF and pseudoscalar
mass determinations of the critical value of $\kappa_{sym}$,
$\kappa_{sym;c}$. The simulations reported here have been performed
at $\beta = 5.50$ which gives $a \sim 0.08\,\mbox{fm}$
(taking the scale $r_0 = 0.5\,\mbox{fm}$ and a linear extrapolation
of $r_0/a$ to the chiral limit). We have furthermore checked that
there is a distinct gap between the distribution of (the modulus of)
the lowest eigenvalue and $0$ indicating that the simulations are
stable on present volumes. Further details of the results in this
write-up will be given in \cite{bietenholz09a}.


\section{Non-degenerate quark masses}

\begin{figure}[htb]

\begin{minipage}{0.40\textwidth}

   \epsfxsize=6.50cm
      \epsfbox{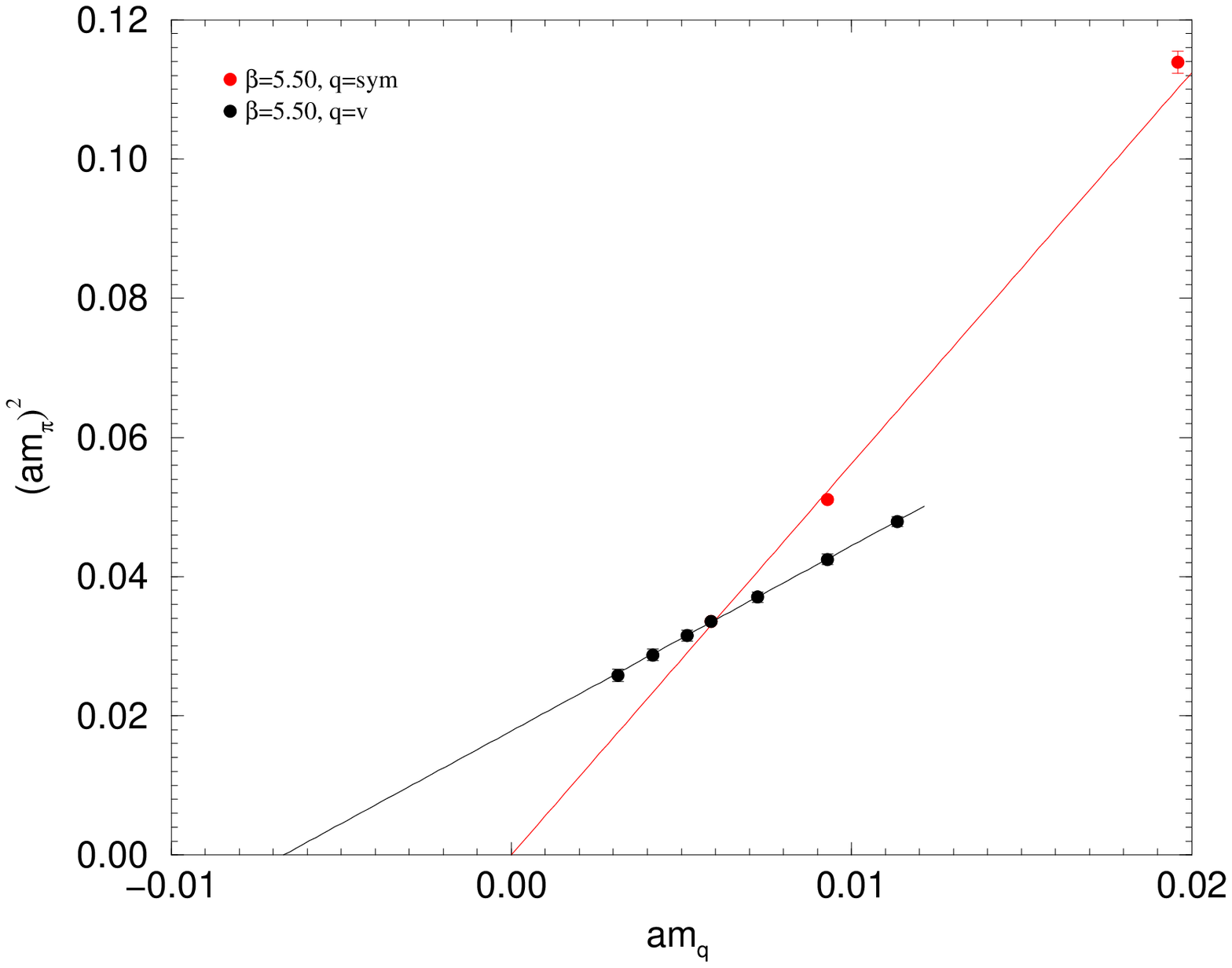}

\end{minipage} \hspace*{0.10\textwidth}
\begin{minipage}{0.40\textwidth}

   \epsfxsize=6.25cm
      \epsfbox{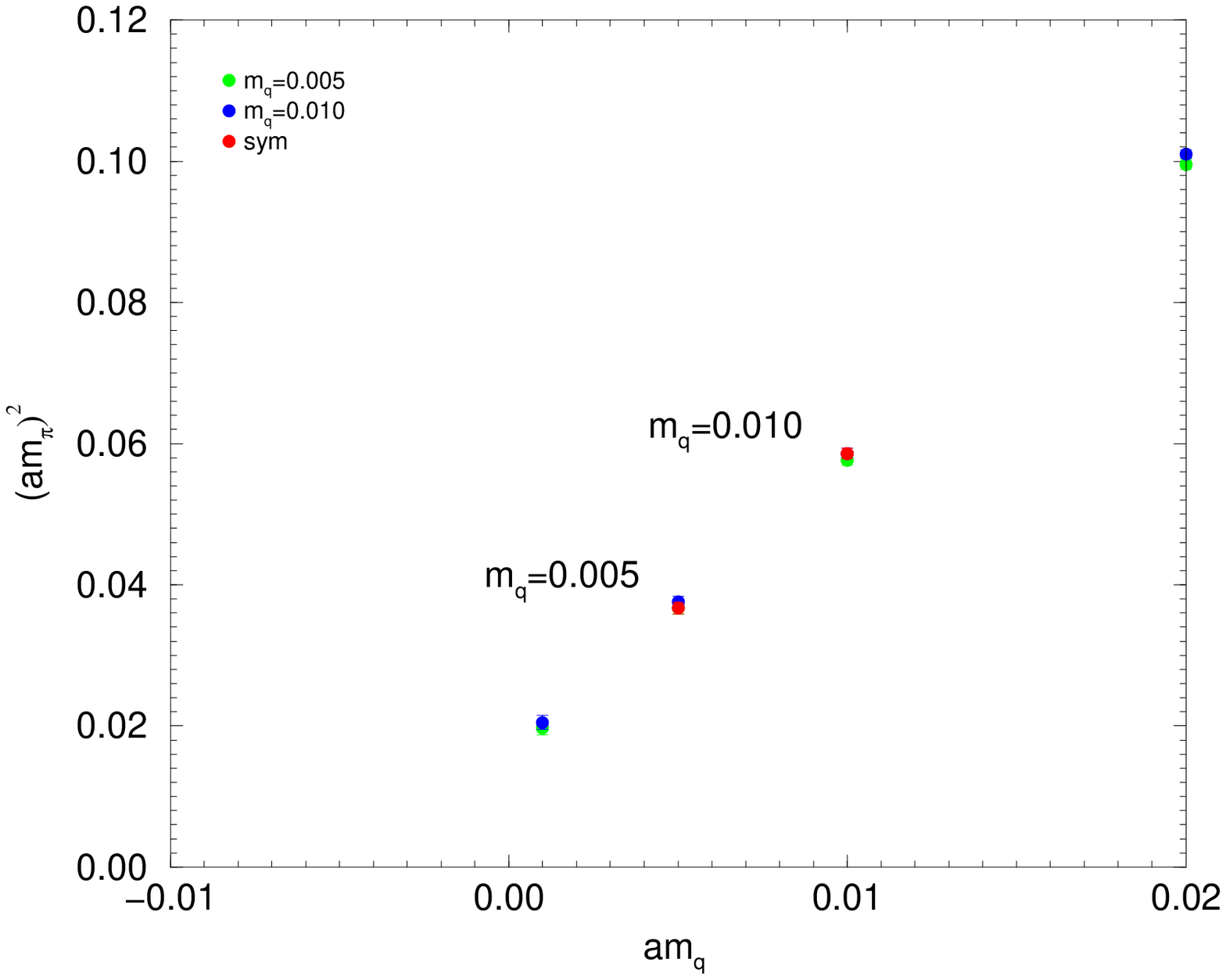}

\end{minipage}

\caption{The left plot shows the $SU_F(3)$ symmetric pseudoscalar
         masses versus $am_q$ (in red) together with the partially
         quenched results (in black) using the mass degenerate
         sea quark $\kappa_{sym} = 0.12090$. (This is the kappa
         value at the crossing point.) The right plot
         shows equivalent domain wall results from \cite{allton08a}
         table V.}

\label{pq_results}
\end{figure}

As mentioned before the problem (at least for clover-like fermions, with no
chiral symmetry) is that singlet and non-singlet quark mass can renormalise
differently,
\begin{eqnarray}
   m^R_q &=& Z_m^{NS}(m_q - \overline{m}) + Z_m^S\overline{m}
                                                                \nonumber \\
         &=& Z_m^{NS}(m_q + \alpha_Z\overline{m}) 
             \quad \mbox{with} \quad
             \alpha_Z = { Z_m^S - Z_m^{NS} \over Z_m^{NS} } \,,
\label{mr2mbare}
\end{eqnarray}
with $q \in \{ sym, l, s, v \}$ (also including possible different
valence quarks to sea quarks) and
\begin{eqnarray}
   \overline{m} = {1 \over 3} (2m_l + m_s) \,.
\end{eqnarray}
(We are assuming here that the quadratic improvement terms,
\cite{bhattacharya05a}, are small, \cite{bietenholz09a}.)
The bare quark mass in eq.~(\ref{mr2mbare}) is defined by
\begin{eqnarray}
   am_q = {1 \over 2} \, 
             \left( { 1 \over \kappa_q} - {1 \over \kappa_{sym;c}} \right) \,.
\end{eqnarray}
Now if $m_{ps}^{q_1q_2}(l,s)$ is the measured pseudoscalar mass (with quarks
$q_1$, $q_2$) then we expect that
\begin{eqnarray}
   (am_K)^2 \equiv (am_{ps}^{q_1q_2})^2 
      \propto am^R_{q_1} + am^R_{q_2}
      \propto am_{q_1} + am_{q_2} + 2\alpha_Z a\overline{m} \,.
\label{ampsK2}
\end{eqnarray}
In particular
\begin{eqnarray}
   (am_\pi)^2 \equiv
   (am_{ps}^{qq})^2 \propto 2(am_q + \alpha_Z a\overline{m}) \,,
\label{amps2}
\end{eqnarray}
which is $\not\propto am_q$ unless $q = l = s = sym$ 
(i.e.\ $m_q = \overline{m}$). That $\alpha_Z$ is non-zero may be easily
seen by considering partially quenched results (i.e.\ results for the 
pseudoscalar mass where the valence quark masses may differ from the
sea quark masses). In Fig.~\ref{pq_results} we show some partially
quenched results and compare the extrapolation with the $SU_F(3)$-symmetric
results. The lines clearly have different gradients. $\alpha_Z$ can be
estimated as $(am_\pi)^2$ vanishes at $\kappa_c^{pq}$ say ($pq$ for
`partially quenched'), giving
\begin{eqnarray}
   \alpha_Z = - { am_q|_{\kappa = \kappa_c^{pq}}\over a\overline{m} }
            =
              { \left( {1 \over \kappa_{sym;c}} - {1 \over \kappa_c^{pq}} \right)
                \over
                \left( {1 \over \kappa_{sym}} - {1 \over \kappa_{sym;c}}
                \right)  } \,,
\end{eqnarray}
(setting $l=s=sym$ in $\overline{m}$). This gives here $\alpha_Z \sim 1.2$
(but the determination is quite sensitive to small changes in $\kappa_{sym;c}$
and $\kappa_c^{pq}$). This is to be compared with domain-wall fermions,
also shown in Fig.~\ref{pq_results}, where the results line up.

\begin{figure}[htb]

\begin{minipage}{0.50\textwidth}

   \epsfxsize=6.00cm \epsfbox{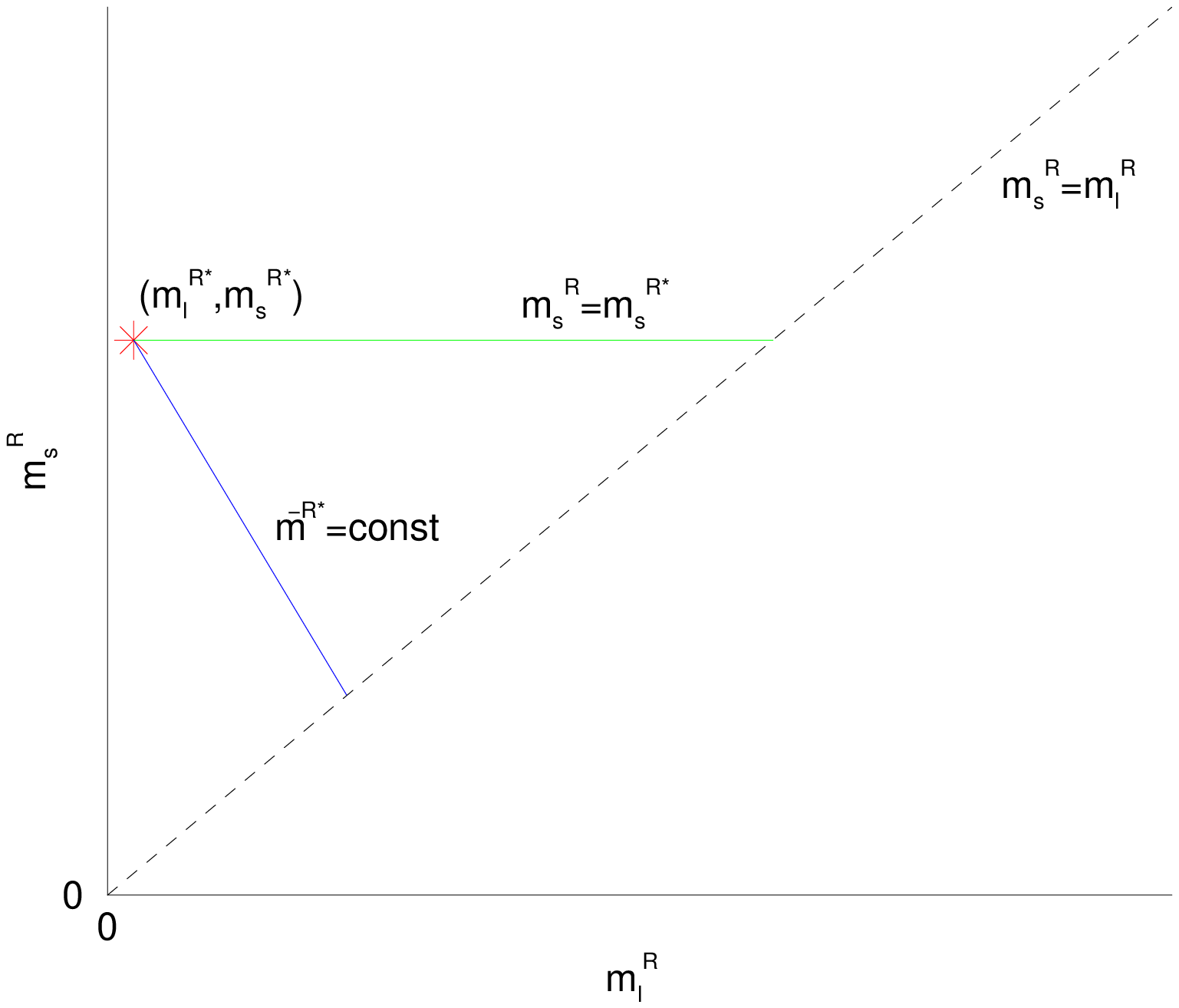}

\end{minipage} \hspace*{0.05\textwidth}
\begin{minipage}{0.30\textwidth}

   \epsfxsize=5.50cm \epsfbox{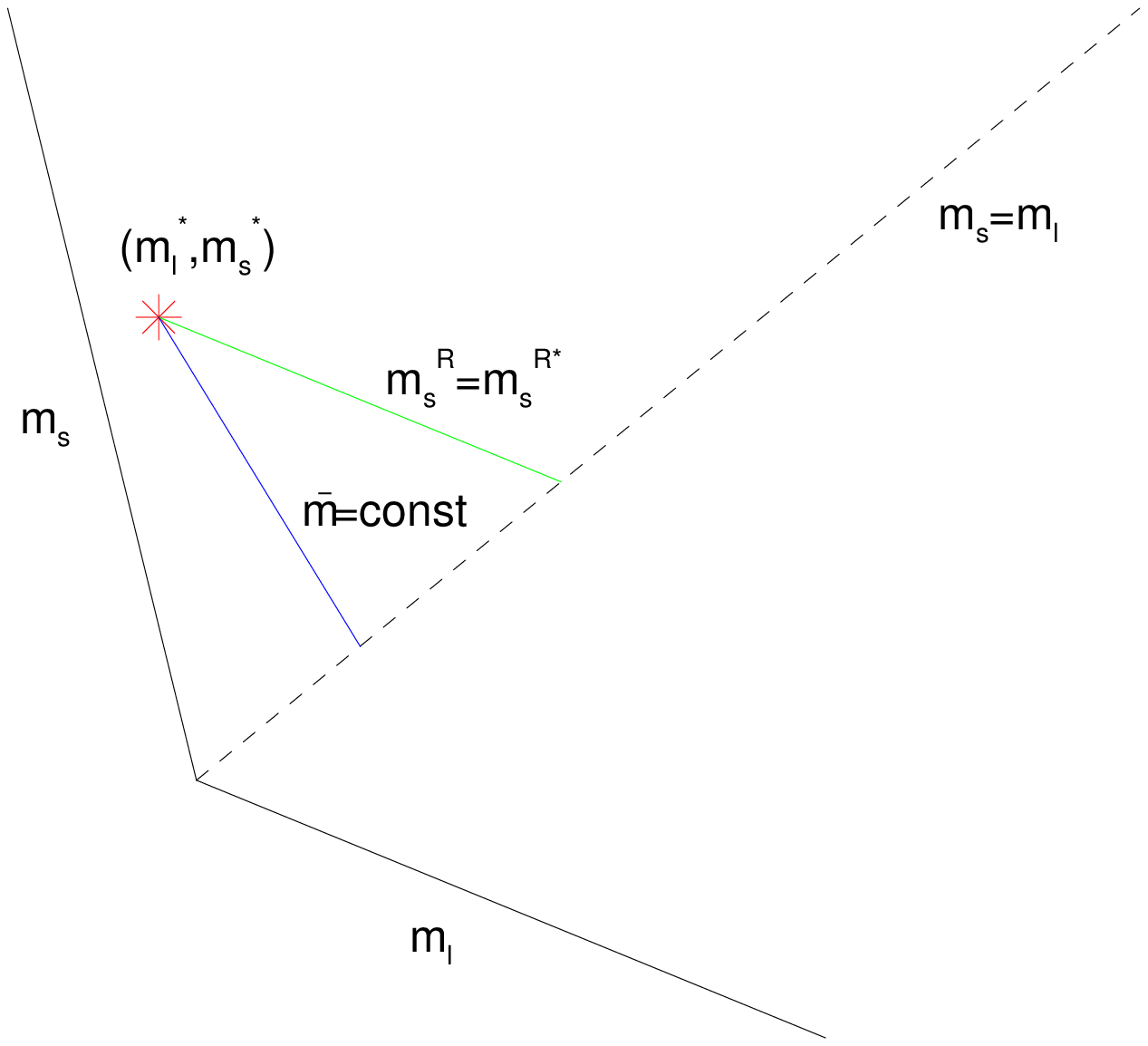}

\end{minipage}

\caption{The left sketch shows the $m_l^R$--$m_s^R$ plane.
         The physical quark masses are denoted by $(m_l^{R*}, m_s^{R*})$.
         The dashed diagonal line is the $SU_F(3)$-symmetric
         line. Two possible paths from this line are shown,
         $m_s^R = m_s^{R*}$ and $\overline{m}^R = \mbox{const.}\,$.
         The right sketch shows the equivalent results in the
         $m_l$--$m_s$ plane.}
\label{sketches}

\end{figure}

We wish to approach the physical point along some path in the
$m_l^R$--$m_s^R$ plane ($m^R_q$ is considered as it is related
to the measurable pseudoscalar mass) from an $SU_F(3)$ symmetric point
($m_l^R = m_s^R \equiv m_{sym}^{R(0)}$). This is depicted in
Fig.~\ref{sketches}.
Two possibilities are $m_s^R = \mbox{const.} = m_s^{R*}$ (i.e.\ strange quark
mass being held constant) or $\overline{m}^R = \mbox{const.}$
(i.e.\ singlet quark mass being held constant).
Note that the region covered is $m_l^R \ge 0$, $m_s^R \ge 0$.
For simulations we need to translate this to unrenormalised
quantities also shown in Fig.~\ref{sketches}. Note that the physical
domain is now
$m_l \ge - ( \third\alpha_Z / (1 + \twothird\alpha_Z) ) m_s$, 
$m_s \ge - ( \twothird\alpha_Z / (1 + \third\alpha_Z) ) m_l$.
While $m_s^R = \mbox{const.} = m_s^{R*}$ translated to bare quark masses
now depends on the difficult-to-determine $\alpha_Z$, the singlet quark
mass $\overline{m}^R \propto \overline{m}$ or
$m_s = (2m^*_l + m^*_s) - 2m_l$ is independent of the value of $\alpha_Z$.
This motivates the choice of this path. As $m_l \searrow m^*_l$
then $m_s \nearrow m^*_s$ i.e.\ the $m_s$--$m_l$ splitting or $m_K$
{\it increases} to its physical value. Other potential advantages include:
the singlet quark mass is correct from the very beginning;
flavour singlet quantities are flat at the symmetric point
-- allowing simpler extrapolations;
numerically the HMC cost change should be moderate along this path.
(These points will be further discussed in \cite{bietenholz09a}.)

\begin{figure}[htb]

\begin{minipage}{0.475\textwidth}

   \epsfxsize=7.00cm \epsfbox{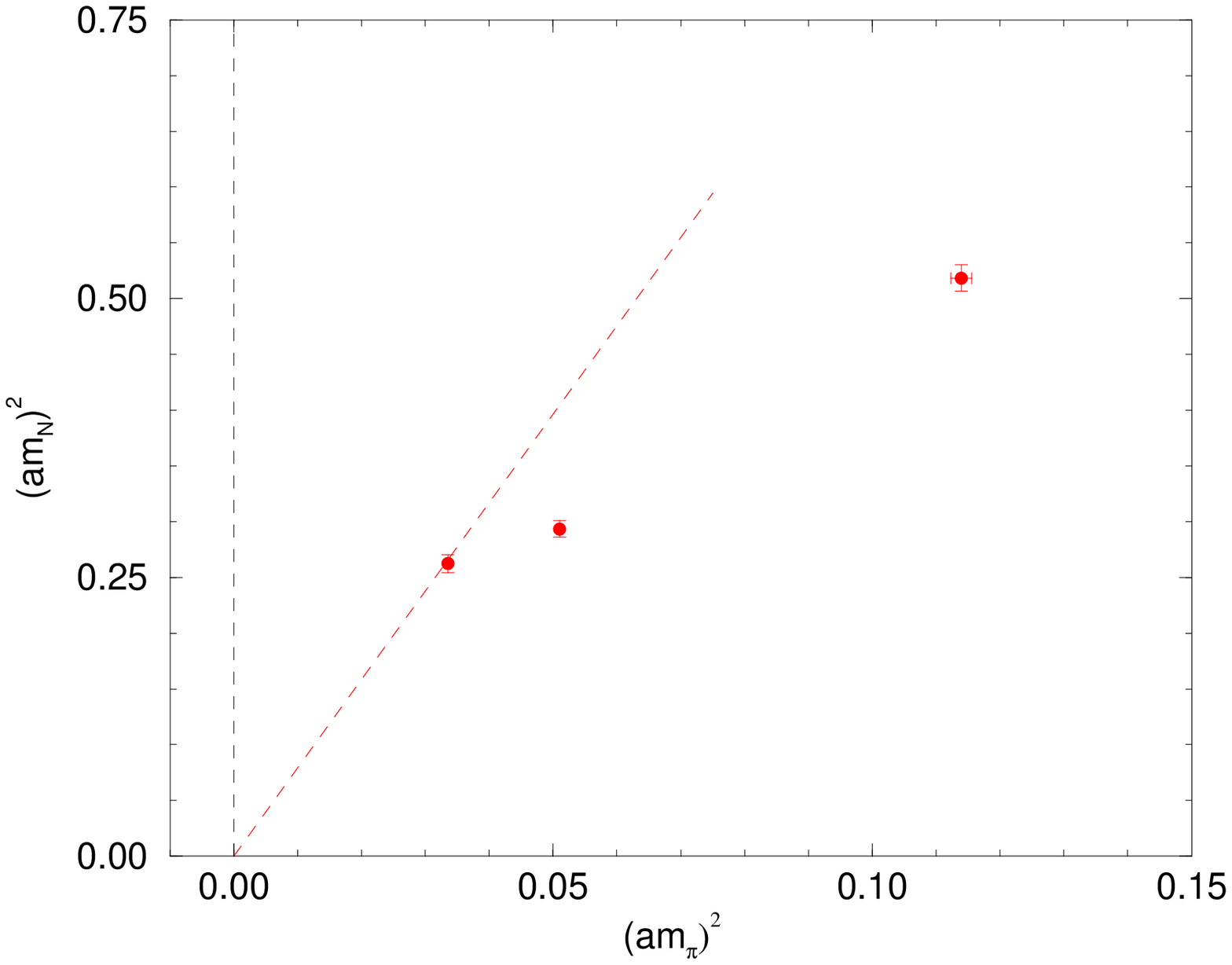}

\end{minipage} \hspace*{0.05\textwidth}
\begin{minipage}{0.475\textwidth}

   \epsfxsize=7.00cm \epsfbox{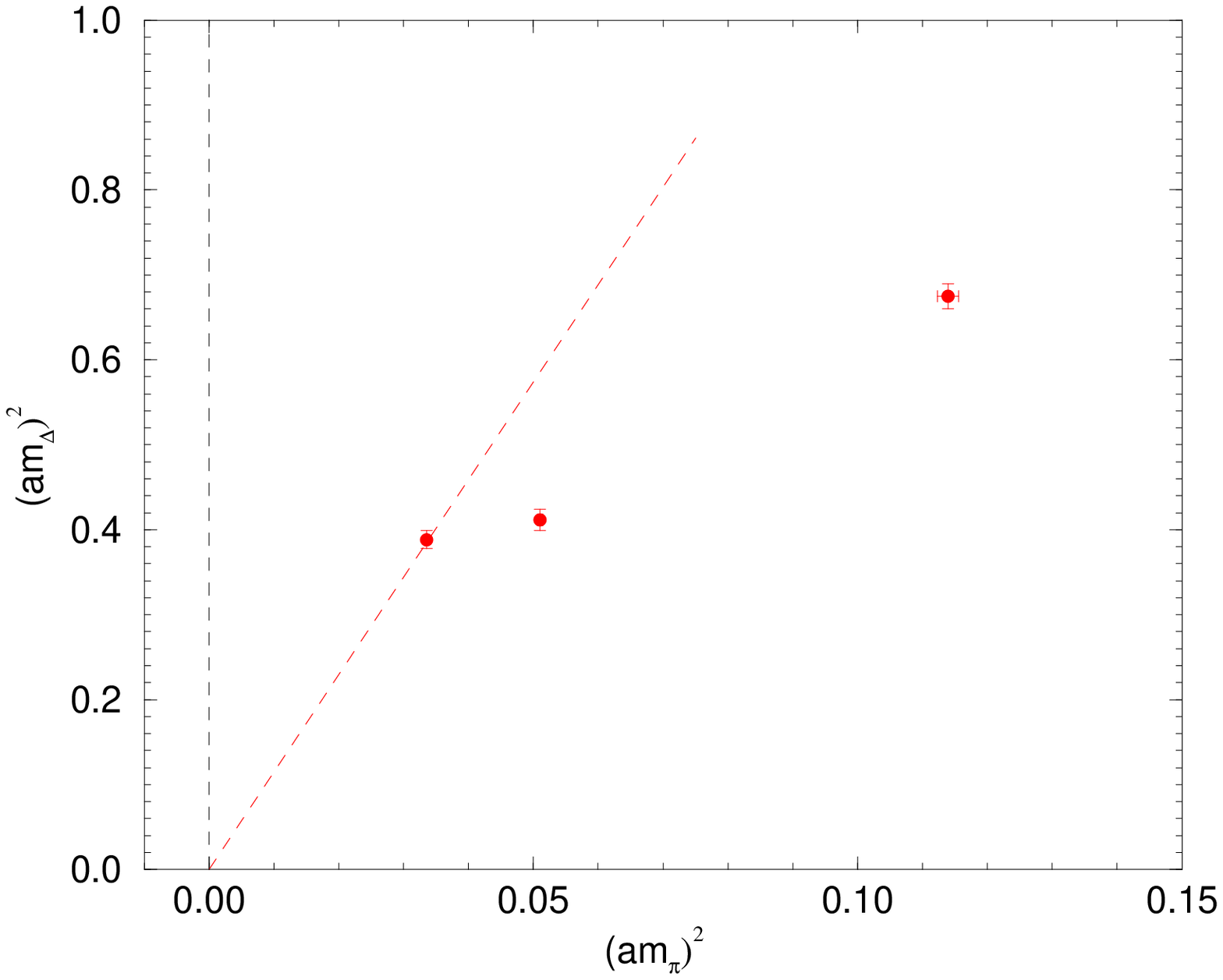}

\end{minipage}

\begin{minipage}{\textwidth}

   \hspace*{1.50in}
   \epsfxsize=7.00cm \epsfbox{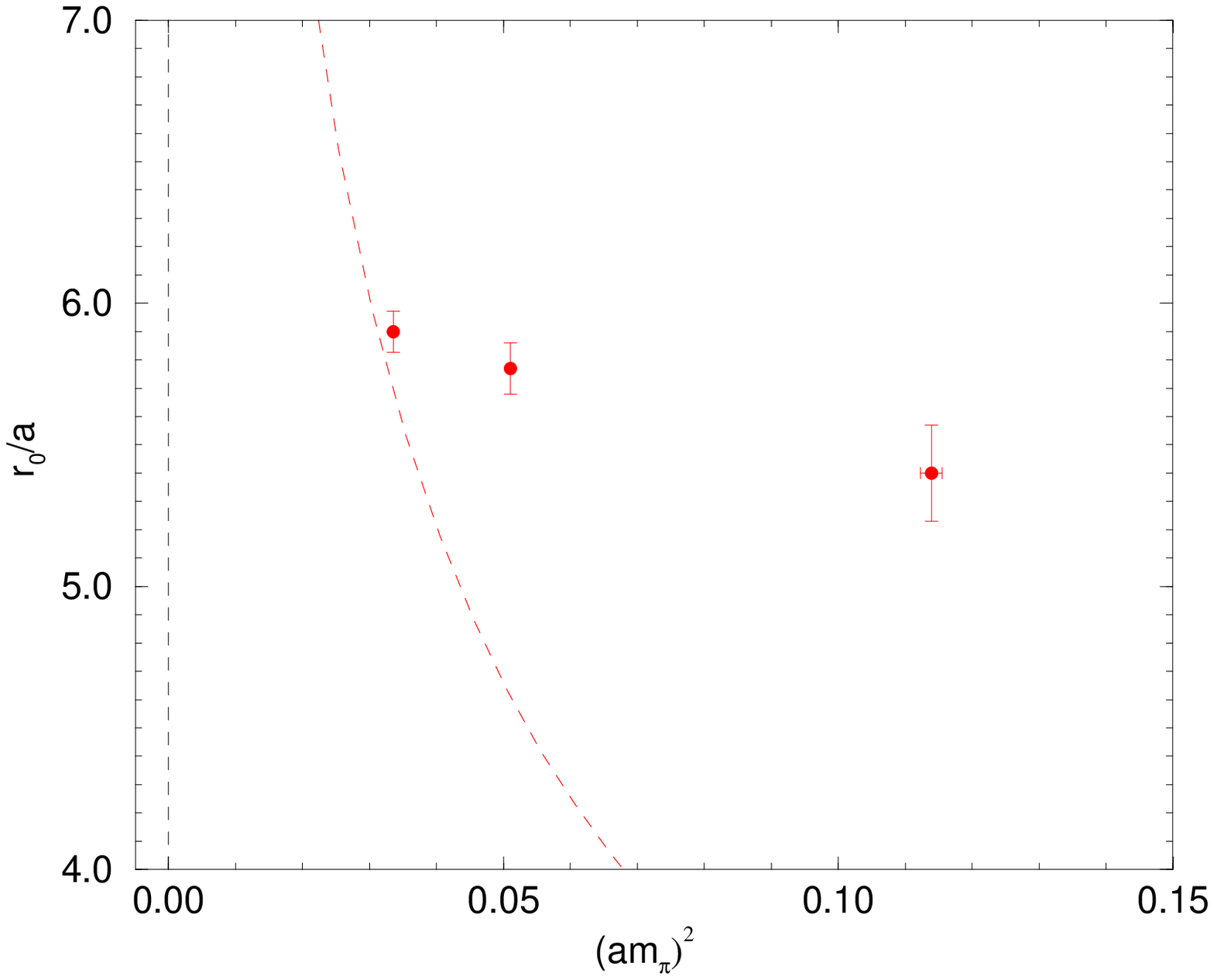}

\end{minipage}

\caption{$(am_N)^2$ against $(am_\pi)^2$ (upper left picture)
         together with the line $(am_N)^2 = 7.93(am_\pi)^2$;
         $(am_\Delta)^2$ against $(am_\pi)^2$ (upper right picture)
         together with the line $(am_\Delta)^2 = 11.5(am_\pi)^2$;
         $r_0/a$ against $(am_\pi)^2$ (lower picture)
         together with the line $(r_0/a)^2 = 1.083/(am_\pi)^2$.
         All data comes from the symmetric points.}

\label{sym_pt_determination}

\end{figure}

Of course practically we must now determine the initial $\kappa_{sym}^{(0)}$
to complete the relation between $\kappa_s$ and $\kappa_l$,
\begin{eqnarray}
  \kappa_s 
      = { 1 \over { {3 \over \kappa_{sym}^{(0)}} - {2 \over \kappa_l} } } \,.
\label{kappas_given_kappal}
\end{eqnarray}
For our path choice, $\kappa_{sym}^{(0)}$ can be implicitly found by
using eqs.~(\ref{ampsK2}), (\ref{amps2}) together with a singlet scale
$X$ (so $X^* = X_{sym}^{(0)}$ here) to relate the known physical point
to the initial symmetric point,
\begin{eqnarray}
   { 1 \over c_X }
      \equiv \left. { {1 \over 3} ( 2m_K^2 + m_\pi^2 ) \over X^2 } \right|^*
      \propto \overline{m}^{R*}
      =  \overline{m}^{R(0)}_{sym}
      \propto { (am_\pi^{(0)})^2 \over (aX_{sym}^{(0)})^2 } \,.
\label{physical_point}
\end{eqnarray}
For $X$ we have several choices. For example
\begin{itemize}

   \item The centre of $\mbox{mass}^2$ of the octet baryons:
         $X_N^2 \equiv \third ( m_N^2 + m_\Sigma^2 + m_\Xi^2 )$
         (stable under strong interaction) giving
         $\third (2m_K^2 + m_\pi^2) / \third (m_N^2+m_\Sigma^2+m_\Xi^2)|^*
          = 0.169 / 1.34  = 1 / 7.93 = (am_\pi^{(0)})^2 / (am_N^{(0)})^2$,

   \item Centre of $\mbox{mass}^2$ of the decuplet baryons:
         $X_\Delta^2 \equiv \third ( 2m_\Delta^2 + m_\Omega^2 )$
         (which decay under strong interactions) giving
         $\third (2m_K^2 + m_\pi^2) / \third (2m_\Delta^2 + m_\Omega^2)|^*
          = 0.169 / 1.94 = 1 / 11.5 = (am_\pi^{(0)})^2 / (am_\Delta^{(0)})^2$,

   \item A gluonic quantity $X_r^2 \equiv 1 / r_0^2$ so
         $r_0^2(2m_K^2+m_\pi^2)|^* = 0.169/0.395^2 = 1.083
          = ( r_0^{(0)} / a )^2 (am_\pi^{(0)})^2$ ($r_0 = 0.5\,\mbox{fm}$).

\end{itemize}
Thus where the numerically determined $SU_F(3)$--symmetric line crosses
with the line $(aX_{sym})^2 = c_X (am_{\pi})^2$ gives our initial
point. The results are shown in Fig.~\ref{sym_pt_determination}.
We see that they are all consistent around the lightest
pseudoscalar mass, namely for $\kappa_{sym}^{(0)} = 0.12090$ which we
shall take as our starting value. From eq.~(\ref{kappas_given_kappal})
we now have a relation between $\kappa_s$ and $\kappa_l$. After some
experimentation we chose the $\kappa_l$, $\kappa_s$ values given
in Table~\ref{kappa_values}.
\begin{table}[h]
   \begin{center}
      \begin{tabular}{ccc}
         $\kappa_l$ & $\kappa_s$ &               \\
         \hline
         0.12083    & 0.12104    &  $m_l > m_s$  \\
         0.12090    & 0.12090    &  $m_l = m_s$  \\
         0.12095    & 0.12080    &  $m_l < m_s$  \\
         0.12100    & 0.12070    &  $m_l < m_s$  \\
         0.12104    & 0.12062    &  $m_l < m_s$  \\
      \end{tabular}
   \end{center}
\caption{Present $(\kappa_l, \kappa_s)$ values (simulated on $24^3\times 48$
         lattices).}
\label{kappa_values}
\end{table}
Note that is possible to choose $\kappa_l$, $\kappa_s$ values
(here $(0.12083, 0.12104)$) such that $m_l > m_s$. In  this strange
world we would expect to see an {\it inversion} of the particle spectrum,
with, for example, the nucleon being the heaviest octet particle.


\section{Hadron spectrum}

\begin{figure}[htb]

\begin{minipage}{0.475\textwidth}

   \epsfxsize=7.00cm \epsfbox{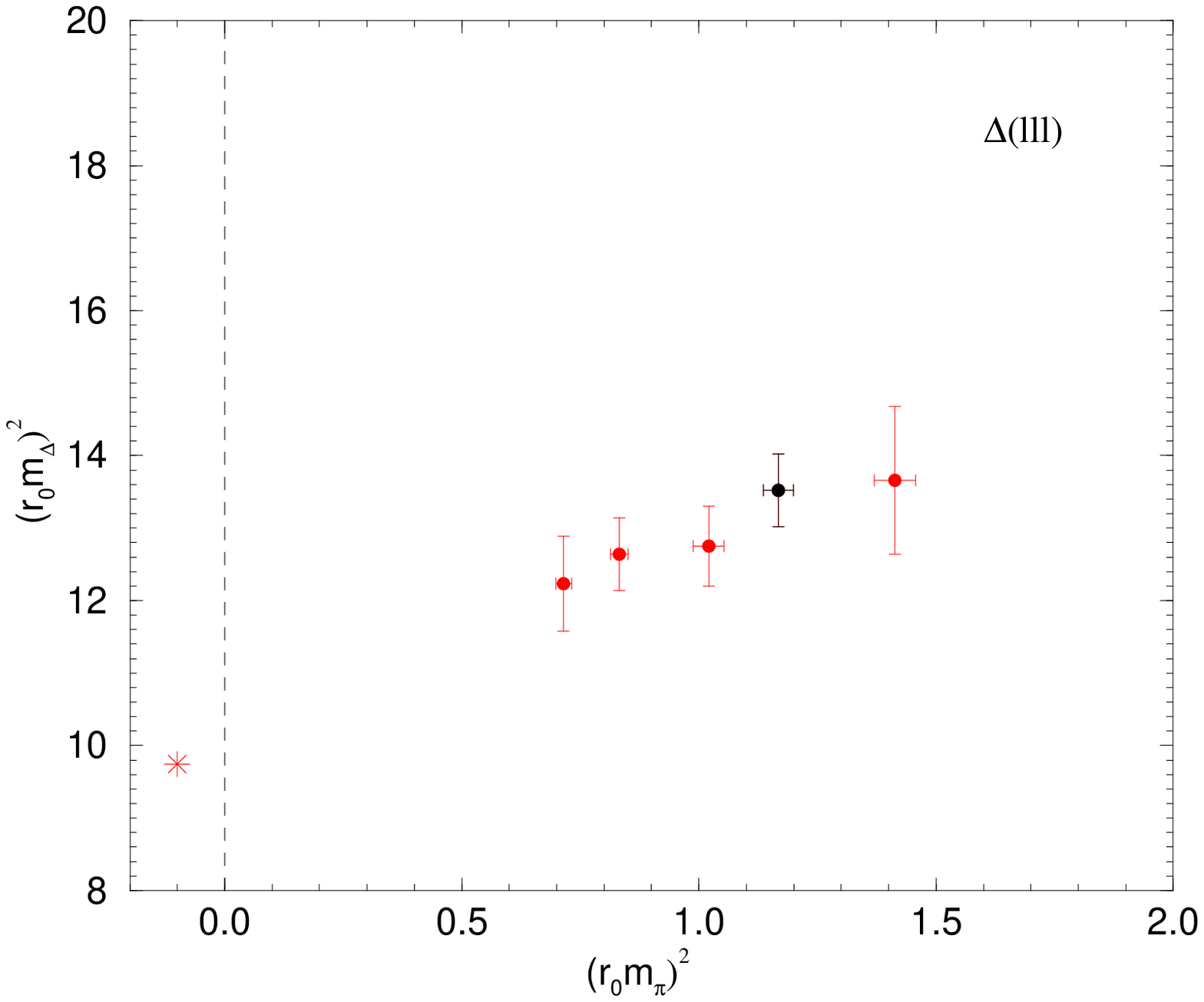}

\end{minipage} \hspace*{0.05\textwidth}
\begin{minipage}{0.475\textwidth}

   \epsfxsize=7.00cm \epsfbox{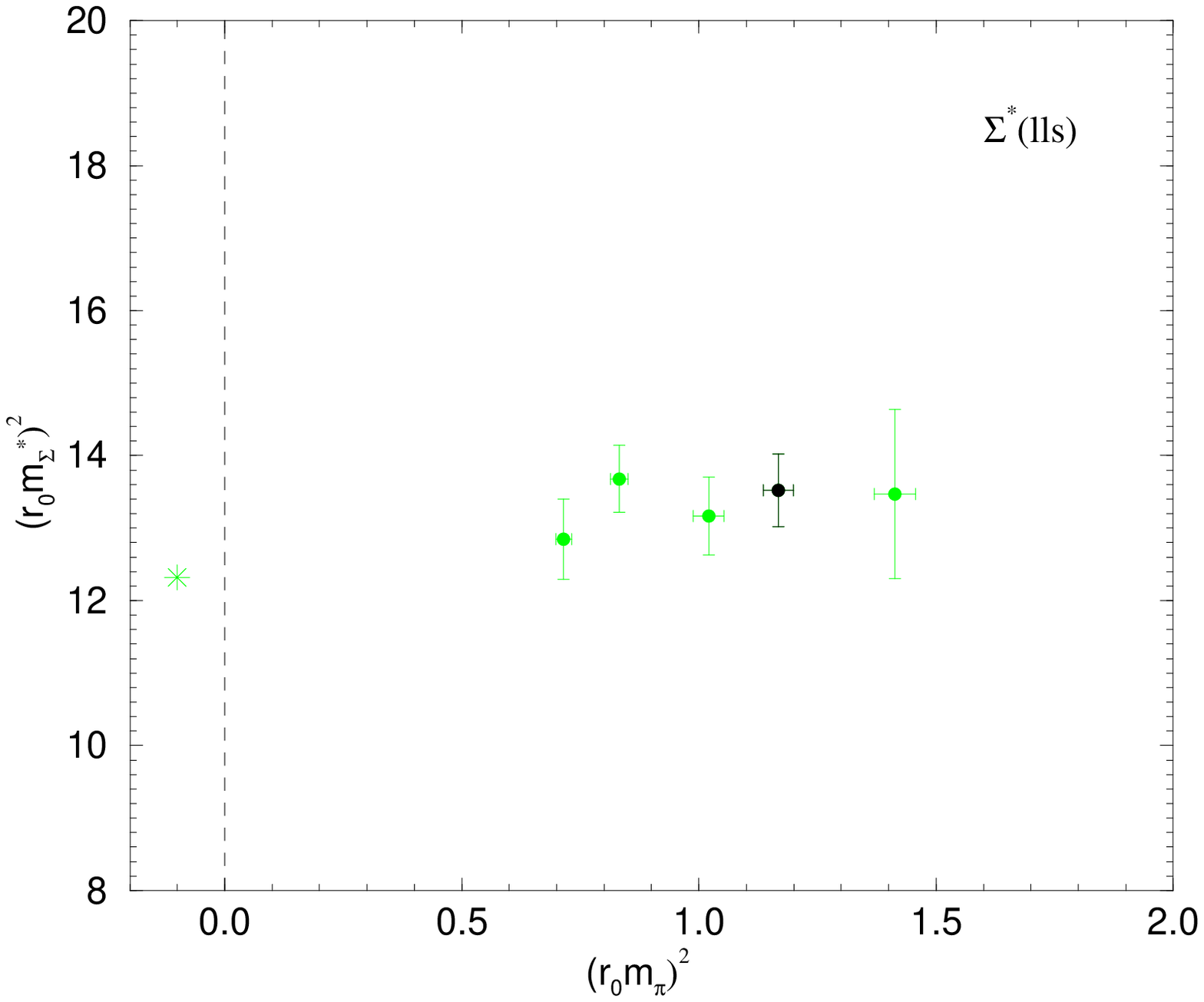}

\end{minipage}

\begin{minipage}{0.475\textwidth}

   \epsfxsize=7.00cm \epsfbox{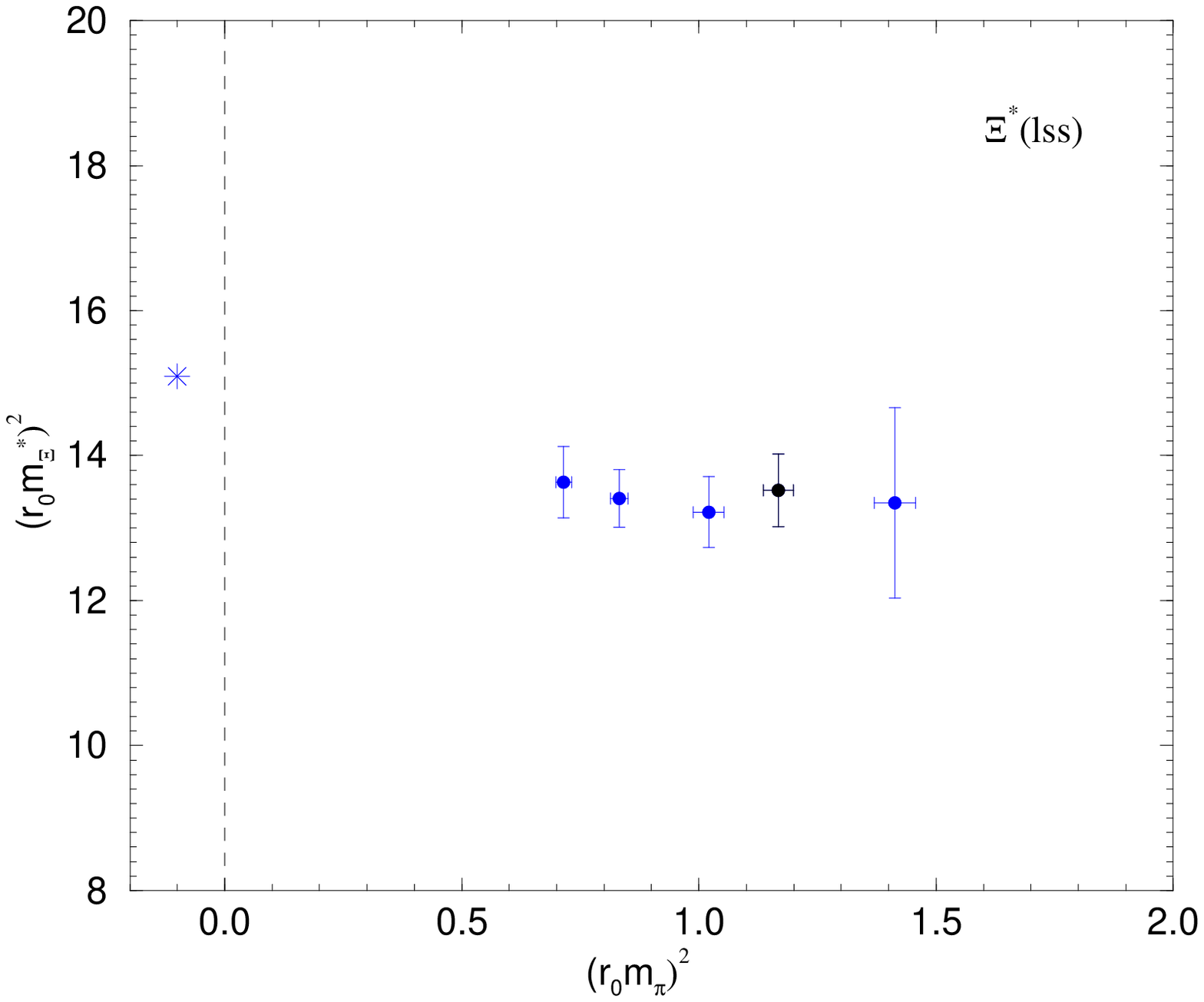}

\end{minipage} \hspace*{0.05\textwidth}
\begin{minipage}{0.475\textwidth}

   \epsfxsize=7.00cm \epsfbox{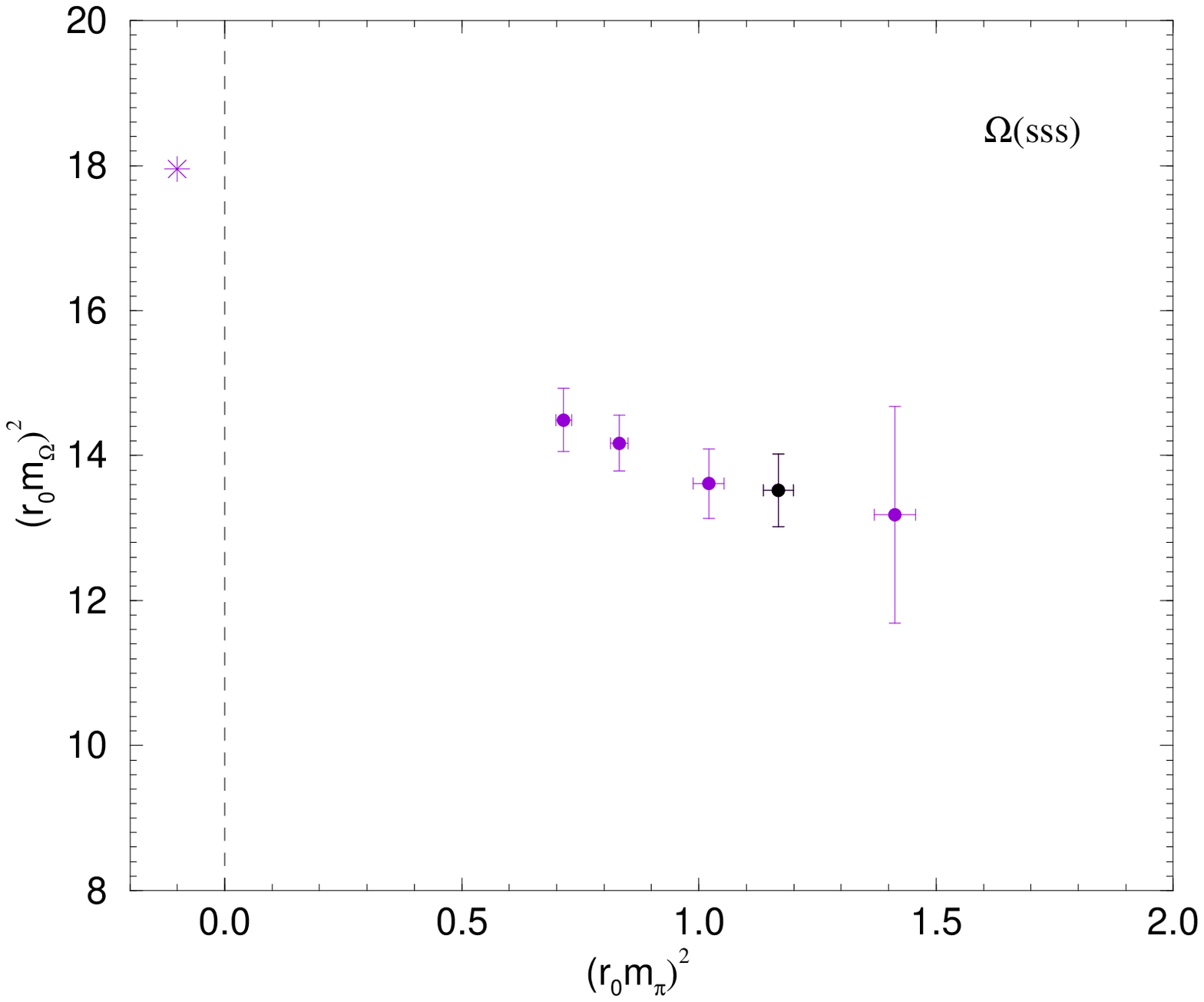}

\end{minipage}

\caption{The decuplet baryons. The upper left picture shows the $\Delta$,
         upper right $\Sigma^*$, lower left $\Xi^*$
         and lower right the $\Omega$.
         The common $SU_F(3)$ symmetric quark mass values are
         shown in black. The experimental values are shown
         with stars.}

\label{hadron_spectrum}

\end{figure}

As an example we now give some results in Fig.~\ref{hadron_spectrum}
for the baryon decuplet: $\Omega$, $\Xi^*$, $\Sigma^*$, $\Delta$.
These are at present low statistics results and are meant to
illustrate the `proof of concept' only.
Noteworthy is that the correct ordering of the particle spectrum
has been achieved.


\section{Conclusions}


For the NP $O(a)$-improved $2+1$ flavour clover action discussed here
we have first found that there is consistency for the critical
$\kappa_{sym}$ between the SF approach and the determination
from the vanishing of the hadron pseudoscalar mass. The quark mass
renormalisation suggests that the simplest way of approaching
the physical point is to hold the singlet quark mass fixed.
Exploratory results for the hadron mass spectrum give splittings
in the correct order (including {\it inversion} when
$m_l > m_s$ i.e.\ we can simulate a strange world where, for example,
the nucleon can decay). Further results will be published elsewhere,
\cite{bietenholz09a}.


\section*{Acknowledgements}


The numerical calculations have been performed on the IBM
BlueGeneL at EPCC (Edinburgh, UK), the BlueGeneL and P at
NIC (J\"ulich, Germany), the SGI ICE 8200 at HLRN (Berlin-Hannover, Germany),
the SGI Altix 4700 at LRZ (Munich, Germany) and JSCC (Moscow, Russia).
We thank all institutions.
The BlueGene codes were optimised using Bagel, \cite{boyle05a}.
This work has been supported in part by
the EU Integrated Infrastructure Initiative Hadron Physics (I3HP) under
contract RII3-CT-2004-506078 and by the DFG under contracts
FOR 465 (Forschergruppe Gitter-Hadronen-Ph\"anomenologie) and
SFB/TR 55 (Hadron Physics from Lattice QCD).



\end{document}